\documentclass[aps,prl,twocolumn,showpacs,groupedaddress]{revtex4}

\usepackage{amsmath,epsfig,graphicx}

\begin{document}

\title{Using A Quantum Dot as a High Frequency Shot Noise Detector}

\author{E. Onac} \altaffiliation{E-mail address: eugen.onac@philips.com}

\author{F. Balestro} \altaffiliation{Present address: Laboratoire Louis
Néel, associ\'e au CNRS, BP 166, F-38042 Grenoble Cedex 9, France}

\author{L. H. Willems van Beveren}

\author{U. Hartmann}\altaffiliation{Physics Department, ASC, and CeNS,
Ludwig-Maximilians-Universit\"{a}t, D-80333 M\"{u}nchen, Germany}

\author{Y. V. Nazarov}
\author{L. P. Kouwenhoven}

\affiliation{Kavli Institute of Nanoscience Delft, Delft University
of Technology, P.O. Box 5046, 2600 GA Delft, The Netherlands}

\begin{abstract}

We present the experimental realization of a Quantum Dot (QD)
operating as a high-frequency noise detector. Current fluctuations
produced in a nearby Quantum Point Contact (QPC) ionize the QD and
induce transport through excited states. The resulting transient
current through the QD represents our detector signal. We
investigate its dependence on the QPC transmission and voltage bias.
We observe and explain a quantum threshold feature and a saturation
in the detector signal. This experimental and theoretical study is
relevant in understanding the backaction of a QPC used as a charge
detector.

\end{abstract}

\pacs{72.70.+m, 73.21.La, 73.23.-b}

\maketitle

On chip noise detection schemes, where device and detector are
capacitively coupled within sub-millimeter length scales, can
benefit from large frequency bandwidths. This results in a good
sensitivity and allows one to study the quantum limit of noise,
where an asymmetry can occur in the spectrum between positive and
negative frequencies. The asymmetry, caused by the difference in the
occurrence probability of emission and absorption processes, can be
probed using \emph{quantum detectors} \cite{Aguado}. In this Letter,
we investigate the transport through a QD under the influence of
high-frequency irradiation generated by a nearby QPC. The QPC
current fluctuations induce photo-ionization , taking the QD out of
Coulomb blockade, thereby allowing sequential tunneling through an
excited state \cite{Oosterkamp,Fujisawa}. By studying this transient
current while changing the QPC parameters, we show that we can
perform high-frequency shot noise detection in the $20$ to $250$ GHz
frequency range.

The granularity of the electron and the stochastic nature of their
transport lead to unavoidable temporal fluctuations in the
electrical current, i.e shot noise \cite{Blanter}. For systems where
transport is completely uncorrelated, like vacuum diodes
\cite{Schottky}, noise is characterized by a Poissonian value of the
power spectral density, $S_I=2eI_{dc}$. Here we use the QPC as a
well-known noise source. When the QPC is driven out of equilibrium,
i.e. by applying an electrochemical potential difference between the
source and the drain of the QPC, a net current will flow if the QPC
is not pinched off. At zero temperature ($k_BT\ll eV_{QPC}$) the
stream of incident electrons is noiseless and shot noise, due to
particle partition, dominates. The electrons are either transmitted
or reflected, with a probability depending on the QPC transmission
$T$. The power density can be written as $S_I=2eI_{dc}F$, where
$F=\sum_{i=0}^N{T_i(1-T_i)}/\sum_{i=0}^N{T_i}$ is the Fano factor
and the summation is over transport channels with transmissions
$T_i$. In this case correlations in the transport are introduced by
the Pauli exclusion principle, resulting in a suppression of noise
below the Poissonian value. Thus, shot noise vanishes if all the
$1D$ quantum channels either fully transmit ($T_i=1$) or reflect
($T_i=0$) \cite{QPCnoise}.

In many recent experiments QPC's are used as charge detectors
\cite{Field}. In this context our experiment provides information
regarding the backaction \cite{Gurvitz,Buks,Avinun} of the QPC when
used as an electrometer for QD devices.

\begin{figure}
\centerline{\epsfig{file=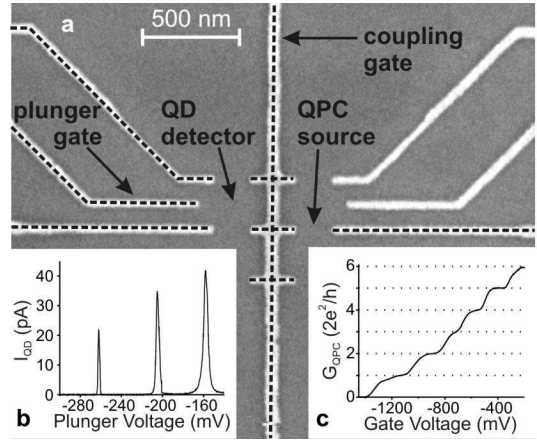, width=7cm,clip=true}}
\caption{\textbf{(a)} Scanning electron micrograph picture of the
gate structure defined on top of the semiconductor heterostructure.
The gates highlighted by dashed lines are used in the present
experiment to define a QD on the left and a QPC on the right. All
other gates are grounded. \textbf{(b)} QD current, $I_{QD}$, versus
plunger gate voltage, for a voltage bias $V_{QD}=30$ $\mu$V, at
$B=1.35$ T. \textbf{(c)} QPC conductance, $G_{QPC}$, as a function
of the gate voltage at $B=0$ T. The potential applied to the
coupling gate is kept constant. The QPC is used as a noise generator
and the QD as a detector.}
 \label{Figure1}
\end{figure}

The QD and the QPC are defined in a GaAs/AlGaAs heterostructure,
containing a two-dimensional electron gas (2DEG) at 90 nm below the
surface, with an electron density \mbox{$n_s = 2.9 \times{10^{11}}$
cm$^{-2}$}. We apply appropriate gate voltages such that we form a
QD on the left and a QPC on the right (Fig.~\ref{Figure1}a). The
lithographic size of the QD is about $250\times{250}$ nm$^2$ and its
charging energy, determined from standard Coulomb blockade
measurements (Fig.~\ref{Figure1}b), is $E_C=1.3$ meV. The QPC
manifests conductance quantization \cite{vanwees}
(Fig.~\ref{Figure1}c), understood in terms of the Landauer formula
$G_{QPC}=(2e^2/h)\sum_{i=1}^NT_i$.

We use the QPC as a noise generator that can be 'switched' ON or OFF
by applying a voltage bias $V_{QPC}$ and/or changing the QPC
transmissions $T_i$. We measure transport through the QD, as a
function of the plunger gate voltage, under the influence of the QPC
noise. A magnetic field perpendicular to the 2DEG can be used to
adjust the transimpedance between the QPC and the QD~
\cite{mag_field}. Stray capacitances in the leads act as short
circuits for high-frequency signals and we use the impedance of edge
states as an insulation between the source-detector part and the
ground of the leads. In this way, the magnetic field enhances the
coupling between the source and the detector. In this Letter we only
present measurements performed using the configuration of
Fig.~\ref{Figure1}a. Measurements with the opposite configuration,
i.e. defining the QD on the right, and the QPC on the left, have
given identical results. Data was taken in a dilution refrigerator,
with an effective electron temperature of $200$ mK.

\begin{figure}
\centerline{\epsfig{file=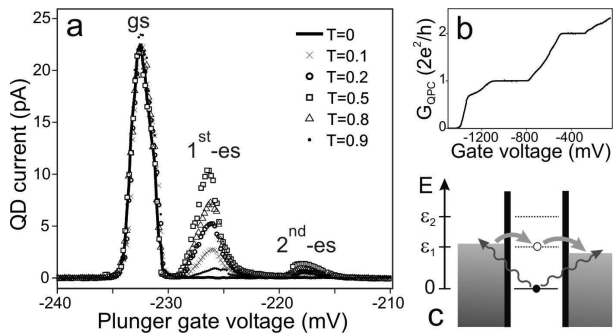, width=8cm,clip=true}}
\caption{\textbf{(a)} Current through the QD, as a function of the
plunger gate voltage, under the influence of shot noise generated by
the QPC. Measurements are performed at \mbox{$B=1.35$ T}, with
$V_{QD}=30$ $\mu$V, $V_{QPC}=1.27$ mV and for different QPC
transmissions. \textbf{(b)} QPC conductance versus the gate voltage
at \mbox{$B=1.35$ T}. \textbf{(c)} Schematic representation of
processes that lead to transport through the excited states of the
QD.} \label{Figure2}
\end{figure}

Our measurements are performed on a QD containing $10$ electrons.
This number was measured using the QPC as a charge detector for the
QD \cite{Field}. The QD voltage bias $V_{QD}=30$ $\mu$V is kept much
smaller than the level spacing ($>200$ $\mu$eV) between the ground
state and the excited states of the QD. When the high-frequency
noise generator is 'switched' OFF (i.e. $V_{QPC}=0$ or $T=\sum_{i}
T_i$ has an integer value), we measure a single Coulomb peak in the
current due to resonant tunneling through the ground state of the QD
(see Fig.~\ref{Figure1}b or Fig.~\ref{Figure2}a for $T=0$). In this
situation, current can only flow through the QD when a charge state
is positioned between the Fermi energies of the leads.

However, if the noise generator is 'switched' ON (i.e. when the QPC
is set out of equilibrium by applying a bias voltage), additional
current peaks emerge in the Coulomb blockade region. The amplitude
of these peaks (labelled $1^{st}$-es and $2^{nd}$-es in
Fig.~\ref{Figure2}a) depends on the QPC transmission and on the
voltage applied to the QPC (Fig.~\ref{Figure4}a). Note that we also
measure this effect when the QPC is current biased.

The additional peaks in the Coulomb blockade regime correspond, in
energy, to the excited states of the QD. These energies are
determined from spectroscopy measurements using large QD voltage
bias. The energy levels of the $1^{st}$ and $2^{nd}$ excited state
relative to the ground state (see Fig.~\ref{Figure2}c) are equal to
$\varepsilon_{1}=245$ $\mu$eV, \mbox{$\varepsilon_{2}=580$ $\mu$eV}
respectively. The QPC gate voltage is adjusted during the QD
measurement in order to compensate for the capacitive coupling of
the plunger gate to the QPC. This allows us to have a fixed
transmission $T$, while measuring the QD. We detect the current
through the $1^{st}$ excited state of the QD in dependence of the
total QPC transmission from $0$ to $2$, and through the $2^{nd}$
excited state for $0<T<1$.

These data can be explained as follows. In the absence of noise,
transport through the excited state is blocked since Coulomb
blockade prevents having electrons in both the ground state and the
excited state simultaneously. The appearance of transport peaks in
the Coulomb blockade region is due to a photo-ionization process
induced by the high-frequency shot noise generated by the QPC
\cite{Oosterkamp2}. Here, an electron in the ground state absorbs
enough energy such that it can leave the dot to either one of the
two leads. Subsequently, a transient current flows through the
excited state, as long as the ground state stays empty
(Fig.~\ref{Figure2}c). This results in the \mbox{appearance of}
conductance peaks, whenever an excited state is aligned between the
Fermi levels of the leads. Current fluctuations through the QPC are
thus converted into a DC current flowing through the excited state
of the QD. This transient current can be analyzed in order to obtain
information regarding the high-frequency fluctuations in the QPC.

For a theoretical description of our results, we first address the
question how the noise couples to the QD? The conversion of QPC
current fluctuations into voltage fluctuations on the QD side is
described by a circuit transimpedance \cite{Aguado} defined as
$\left| Z(\omega) \right|=\sqrt{{S_V(\omega)}/{S_I(\omega)}}$, with
$S_I(\omega)$ the current spectral density of noise generated by the
QPC and $S_V(\omega)$ the power spectral density of voltage
fluctuations at one barrier of the QD. This can be expressed as
$\left| Z(\omega) \right| \approx \left| Z(0) \right| =\kappa R_K$,
where $R_K=e^2/h=25.8$ k$\Omega$ is the quantum resistance and
$\kappa$ is a dimensionless parameter describing the coupling
between different QPC channels and QD barriers. In our model, we
define four different $\kappa$ coefficients depending on the channel
involved in the QPC, and the barrier of the QD: $\kappa_{L,1}$ and
$\kappa_{L,2}$ are the coupling coefficients between the first
respectively the second channel of the QPC and the left barrier of
the QD, and $\kappa_{R,1}$ and $\kappa_{R,2}$ describe the coupling
of the QPC channels to the right barrier. Experimentally, we can
adjust the QD barriers in order to have symmetric escape rates to
the left and the right reservoirs. The absence of pumping effects
(i.e. regions where $I_{QD}<0$) close to the Coulomb peaks (see
Fig.~\ref{Figure2}a) indicates symmetric coupling for the QD
barriers: $\kappa_{R}=\kappa_{L}$. The only independent coupling
parameters are $\kappa_{1}\neq\kappa_{2}$ corresponding to the first
two QPC channels. As already discussed, a perpendicular magnetic
field can be used to increase the coupling parameter $\kappa$
(although this is not understood quantitatively).

The second question we address is what kind of energies and cut-off
frequencies are involved in the photo-ionization process? In the low
temperature limit, two energy scales are important for the detection
mechanism. First, the energy difference $\varepsilon$ between the
ground and the excited state of the QD is relevant, as the
photo-ionization process pumps an electron out from the ground
state. This level spacing (see Fig.~\ref{Figure2}c) sets a detector
cut-off frequency $\nu_{QD}=\varepsilon/h$, representing the minimum
frequency that can induce photo-ionization (the minimum energy that
can be detected, assuming single photon PAT processes). The second
relevant energy is provided by the QPC bias. This gives the cut-off
frequency for the noise generator $\nu_{QPC}=eV_{QPC}/h$,
corresponding to the maximum frequency that can be emitted (for
independent tunneling events in the QPC). Thus, the frequencies
contributing to the PAT process are in the range
[$\nu_{QD}$,$\nu_{QPC}$]. For the measurements in
Fig.~\ref{Figure2}a, $V_{QPC}=1.27$ mV, which corresponds to
$\nu_{QPC}=317$ GHz, and, depending on the $1^{st}$ or $2^{nd}$
excited states, $\nu_{QD}$ is equal to $\varepsilon_{1}/h=59$ GHz or
$\varepsilon_{2}/h=140$ GHz. These set two different detection
bandwidths for the $1^{st}$ and the $2^{nd}$ excited state, leading
to different amplitudes for the detector signal (i.e. the peak
height of the transient current).

Our model considers PAT in a QD. Noise generated by the QPC induces
potential fluctuations between the QD energy levels and the
electrochemical potentials in the leads. These fluctuations modify
the tunneling rates $\Gamma_L$, $\Gamma_R$ between the QD and its
source and drain leads. The change can be described using the theory
of energy exchange with the environment \cite{Ingold}. If we
consider the simple case of "classical" shot noise, where the power
spectrum is white, the photo-ionization probability has a Lorenztian
shape $P_i(E)=\frac{1}{\pi{w_i}}\frac{1}{1+{E^2}/{w_i^2}}$. The
width $w_i=8\pi^2\kappa_i^2T(1-T)eV_{QPC}$ \cite{Aguado} includes
the coupling coefficient as well as the "classical" noise power
emitted by the QPC.
\begin{figure}
\centerline{\epsfig{file=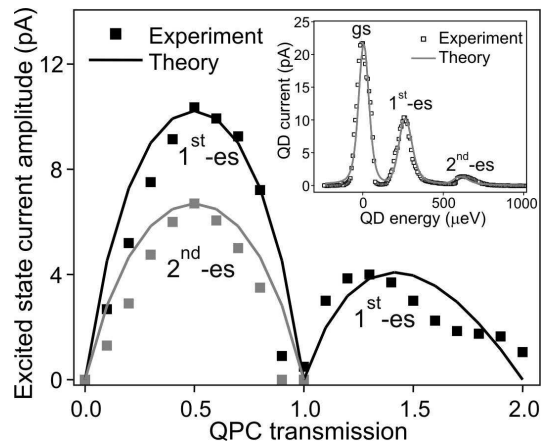, width=7cm,clip=true}}
\caption{Amplitude of the current through the excited state of the
QD versus QPC transmission. Measurements are performed at $1.35$
Tesla with $V_{QD}=30$ $\mu$V and $V_{QPC}=1.27$ mV. The current
amplitude through the $2^{nd}$ excited state for $0<T<1$, and
through the $1^{st}$ excited state for $1<T<2$ have been multiplied
by a factor of 5 for clarity. Inset: QD current as a function of the
QD energy for a QPC transmission $T=0.5$. Experimental points are in
good agreement with the solid theoretical curve. The plunger gate
value is converted in QD energy for clarity.}
  \label{Figure3}
\end{figure}

Using this theoretical model, we can fit the experimental results
and obtain the parameters that characterize our system. We first
extract the tunneling rate through the ground state of the QD by
fitting the Coulomb peak when the noise generator is 'switched' OFF
(no additional peaks in the Coulomb blockade regime). We tune the
system, by applying appropriate gate voltages on the electrodes, in
order to have a symmetric QD: the two tunneling rates from QD to
source ($\Gamma_L$) and drain ($\Gamma_R$) are equal. From the fit
results a value of $\Gamma_L=\Gamma_R=0.575$ GHz. The electron
temperature, the voltage across the QD and across the QPC are known
parameters, and are respectively equal to $200$ mK, $30$ $\mu$V, and
$1.27$ mV. In order to explain the additional peaks in the Coulomb
blockade regime, and the modulation of these peaks as a function of
the QPC transmission, we introduce the following fit parameters: the
escape rates $\Gamma_1^{es}$ and $\Gamma_2^{es}$ of the first and
the second excited state, the coupling coefficients $\kappa_1$ and
$\kappa_2$ to the first and the second channel of the QPC. By using
this set of four fitting parameters, we are able to obtain a good
fit for the QD current dependence on the plunger gate voltage, in
the presence of noise (see inset to Fig.~\ref{Figure3}). The
resulting fit values are reasonable: $\Gamma_1^{es}=5.75$ GHz,
$\Gamma_2^{es}=4.03$ GHz, $\kappa_1=1.67\times10^{-2}$,
$\kappa_2=4.83\times10^{-3}$. The coupling coefficients are
difficult to estimate and they depend strongly on the details of the
electromagnetic environment (e.g. on the geometry of the sample).
The one order of magnitude difference between the coupling to the
first and the second channel of the QPC is likely due to the
shunting between the leads, provided by the first, conducting
channel.

In Fig.~\ref{Figure3} we plot the current flowing through the
$1^{st}$ (black square) and the $2^{nd}$ (gray square) excited state
of the QD as a function of the QPC transmission $T$. The points
simply represent the current peak values as extracted from
measurements presented in Fig.~\ref{Figure2}a. We find that the QD
detector signal is modulated by changing the QPC transmission in
accordance to the shot noise theory: the noise vanishes for integer
values ($T=1$ or $T=2$) and is maximal for $T=0.5$ and close to
$T=1.4$. The solid lines represent theoretical calculated values by
making use of the previous determined parameters. We note that one
set of fitting parameters can be used to describe the PAT signal
dependence on both the QD energy and the QPC noise power. In
Fig.~\ref{Figure3}, a factor $5$ has been introduced in the vertical
scaling for the $2^{nd}$ excited state, and also for the $1^{st}$
excited state from $T=1$ to $2$, for clarity. The suppression of the
detector signal for these two cases was already discussed: low
amplitude of the $2^{nd}$ excited state current is due to a smaller
detection bandwidth, while the noise generated by the second QPC
channel is partly screened by the electrons flowing through the
first, ballistic channel. Screening is also the reason a QPC is less
sensitive as an electrometer in the $T>1$ transmission range.

\begin{figure}
\centerline{\epsfig{file=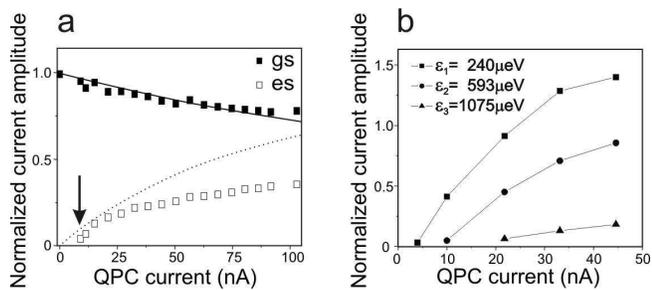, width=8.5cm,clip=true}}
\caption{\textbf{(a)} Normalized amplitude of current flowing
through the ground and excited state as a function of the QPC
current. Measurements (presented as squares) are performed at
$B=2.6$ T, with the QPC current biased at half transmission $T=0.5$.
The solid and the dotted curves represent theoretical values
obtained using parameters from the measurements at $B=1.35$ T.
\textbf{(b)} Quantum limit cut-off frequencies corresponding to
three excited states in the quantum dot detector. QPC transmission
is set to $T\approx 1/7$.} \label{Figure4}
\end{figure}

In Fig.~\ref{Figure4}a, we measure and theoretically compute the
saturation of the excited state current as a function of $I_{QPC}$.
The plot presents the current amplitude for the ground and the
excited state, normalized to the amplitude of the Coulomb peak in
the absence of noise. We clearly see that the amplitude of the
excited state increases as a function of the QPC bias, while the
amplitude of the ground state decreases. For the excited state, the
difference between the experimental and the calculated amplitude, at
large QPC currents ($>25$ nA), is due to a second excited state and
relaxation processes neglected in the theoretical calculation.

A distinct quantum feature present in the experimental measurements
is the existence of a cut-off in the values of the QPC voltage bias
(indicated by the arrow in Fig.~\ref{Figure4}a). This corresponds to
the condition $\nu_{QPC}=\nu_{QD}$ and represents the minimum QPC
voltage bias for which the detection mechanism works. For smaller
bias voltages the emission side of the QPC noise is zero at the
frequencies $\nu>\nu_{QD}$ where the QD detector is sensitive. The
theoretical results are obtained from a "classical", frequency
independent expression for shot noise and, subsequently, they do not
show this cut-off. Noise cut-off frequencies corresponding to three
excited states are measured and presented in Fig.~\ref{Figure4}b. At
higher noise power, we measure a saturation for both amplitudes of
current through the excited and the ground state. This phenomenon
can be understood as reaching an equilibrium between PAT and QD
relaxation processes.

In conclusion, we use a QD as an on chip quantum detector to
achieve, for the first time, very high-frequency (in the range
[$20$-$250$] GHz) shot noise measurements. We measure the cut-off,
determined by the QPC bias, in the emission part of the noise
spectrum. The detection process can also be viewed as a backaction
of the QPC when used as a QD electrometer and represents an
important source for the dark counts in the single-shot readout of
individual electron spins in quantum dots \cite{Elzerman2}.

We are grateful to R. Hanson for helping with the sample
fabrication. We acknowledge technical assistance from R. Schouten
and A. van der Enden. Financial support provided by the Dutch
Foundation for Fundamental Research on Matter (FOM).


\begin{thebibliography}{15}

\bibitem{Aguado}
R. Aguado and L. P. Kouwenhoven, Phys. Rev. Lett. {\bf 84}, 1986
(2000).

\bibitem{Oosterkamp}
T. H. Oosterkamp \emph{et al.}, Nature {\bf395}, 873 (1998).


\bibitem{Fujisawa}
T. Fujisawa, Y. Tokura, and Y. Hirayama, Phys. Rev. B {\bf 63},
081304(R) (2001).

\bibitem{Blanter}
{Y. M. Blanter and M. B\"uttiker, Phys. Rep. {\bf 336}, 1 (2000).}

\bibitem{Schottky}
W. Schottky, Ann. Phys. (Leipzig){\bf 57}, 541 (1918).

\bibitem{QPCnoise}
Y. P. Li \emph{et al.}, Appl. Phys. Lett. {\bf 57}, 774 (1990); M.
Reznikov, M. Heiblum, H. Shtrikman, and D. Mahalu, Phys. Rev. Lett.
{\bf 75}, 3340 (1995); A. Kumar \emph{et al.}, Phys. Rev. Lett. {\bf
76}, 2778 (1996).

\bibitem{Field}
M. Field \emph{et al.}, Phys. Rev. Lett. {\bf 70}, 1311 (1993).



\bibitem{Gurvitz}
S. A. Gurvitz \emph{et al.}, Phys. Rev. Lett. {\bf 91}, 066801
(2003).

\bibitem{Buks}
E. Buks \emph{et al.}, Nature {\bf 391}, 871-874 (1998).

\bibitem{Avinun}
M. Avinun-Kalish \emph{et al.}, Phys. Rev. Lett. {\bf 92}, 156801
(2004).


\bibitem{vanwees}
B. J. van Wees \emph{et al.}, Phys. Rev. Lett. {\bf 60}, 848 (1988).



\bibitem{mag_field}
The precise magnetic field values were chosen corresponding to
minima in the Shubnikov-de Haas oscillations in the bulk 2DEG. This
implies that no back scattering occurs and that the incident stream
of electrons is noiseless [e.g. see M. Henny \emph{et al.}, Science
{\bf 284}, 296 (1999)].

\bibitem{Oosterkamp2}
{T. H. Oosterkamp \emph{et al.}, Phys. Rev. Lett. \textbf{78}, 1536
(1997).}

\bibitem{Ingold}
G. L. Ingold and Y. V. Nazarov, in Single Charge Tunneling, edited
by H. Grabert and M. H. Devoret (Plenum, New York, 1992)



\bibitem{Elzerman2}
{J. M. Elzerman \emph{et al.}, Nature {\bf430}, 431 (2004).}



\end{thebibliography}
\end{document}